\documentclass[nofootinbib,onecolumn,preprint,showpacs,superscriptaddress]{revtex4-1}

\usepackage[utf8]{inputenc}
\usepackage{mathtools}
\usepackage{graphicx}
\usepackage{caption}
\usepackage{subcaption}
\usepackage{xcolor}
\usepackage{soul}
\usepackage{enumerate}
\usepackage{amsmath,amsfonts,amssymb,amsthm}
\usepackage{commath}
\usepackage{indentfirst}
\usepackage{setspace}
\usepackage{dsfont}
\usepackage{lastpage}
\usepackage{float}
\usepackage{epstopdf}
\usepackage{scrextend}
\usepackage{footnote}
\usepackage[ruled,vlined]{algorithm2e}
\usepackage{times}
\usepackage{feynmf}
\usepackage{physics}
\usepackage{cancel}
\usepackage{comment}
\usepackage{subcaption}
\usepackage{booktabs}
\usepackage{mathtools}

\usepackage{hyperref}

\begin{document}

\def\a{\alpha}
\def\b{\beta}
\def\G{\Gamma}
\def\t{\theta}
\def\s{\sigma}
\def\d{\partial}
\def\ddl{\mathcal{L}}
\def\ddlint{\ddl_{\text{int}}}
\def\ddlem{\frac{1}{4}F_{\mu\nu}F^{\mu\nu}}
\def\l{\left}
\def\r{\right}
\def\L{\Lambda}
\def\meio{\frac{1}{2}}
\def\ceb{\coth\left(n \pi \frac{E}{B} \right)}
\def\cbe{\coth\left(n \pi \frac{B}{E} \right)}
\def\mg{\mathcal{G}}
\def\mf{\mathcal{F}}
\def\Msun{M_{\odot}}
\def\Qlim{Q_{\mathrm{lim}}}
\def\ra{\rightarrow}
\def\o{\mathcal{O}}
\def\aring{\stackrel{\circ}}
\def\rmunu{R_{\mu\nu}}\textbf{}
\def\espacinho{\vspace{0.4cm}}
\def\jumph{\hspace{0.2cm}}
\def\jumpheq{\hspace{0.2cm}=\hspace{0.2cm}}
\def\sen{\sin}
\def\checkmark{\tikz\fill[scale=0.4](0,.35) -- (.25,0) -- (1,.7) -- (.25,.15) -- cycle;} 
\def\vp{\vec{p}}
\def\vk{\vec{k}}
\def\vd{\vec{d}}
\def\vx{\vec{x}}
\def\vy{\vec{y}}
\def\vq{\vec{q}}
\def\vr{\vec{r}}
\def\dtq{\frac{d^3q}{(2\pi)^3}}
\def\dtfourk{\frac{d^4k}{(2\pi)^4}}
\def\empty{\hspace{0.15cm}}
\def\psibar{\Bar{\psi}}
\def\rscalar{\mathcal{R}}
\makeatletter
\newcommand{\vast}{\bBigg@{3}}
\newcommand{\Vast}{\bBigg@{4}}
\newcommand{\Vaster}{\bBigg@{4}}

\title{Curvature Corrections to the Yukawa Potential in Tolman Metrics} 

\author{J. V. Zamperlini}
\email{joao.zamperlini@posgrad.ufsc.br}
\affiliation{Departamento de Física, CFM - Universidade Federal de \\ Santa Catarina; C.P. 476, CEP 88.040-900, Florianópolis, SC, Brazil}
\affiliation{Departamento de Física (DAFIS), Universidade Tecnológica Federal do Paraná (UTFPR), Curitiba - PR, 80230-901, Brasil}

\author{C. C. Barros Jr.}
\email{barros.celso@ufsc.br}

\affiliation{Departamento de Física, CFM - Universidade Federal de \\ Santa Catarina; C.P. 476, CEP 88.040-900, Florianópolis, SC, Brazil} 

\begin{abstract}
This work investigates curvature-induced modifications to the Yukawa potential in static, spherically symmetric spacetimes described by Tolman metrics, focusing on their implications for compact stellar objects, with particular application to solutions IV and VI. Motivated by the interplay of quantum interactions and strong gravitational fields in systems like neutron stars, we derive explicit corrections to the Yukawa potential for these metrics based on recent work. \textcolor{black}{Revisiting} the previous result\textcolor{black}{, contrary to what was found, that} curvature corrections break \textcolor{black}{the interacting potential} radial symmetry near a highly charged black hole, we show that Tolman metric \textcolor{black}{corrections still provide the same symmetry in the local inertial frame}. Numerical estimates for astrophysical objects reveal energy shifts of the order of $10^{-34}\ \mathrm{MeV}$ \textcolor{black}{for the solution IV.} The Tolman VI solution, while singular at the center, yields comparable corrections for most of the fluid sphere radius. \textcolor{black}{A detailed analysis of the repulsive or attractive feature of the curvature corrections for a local observer is done for each scenario}. \textcolor{black}{Despite providing small corrections,} these results highlight the role of spacetime geometry in shaping quantum interactions and provide a foundation for future studies of nuclear interactions within the context of relativistic stars.
\end{abstract}

\maketitle

\section{Introduction}

In recent decades, reconciling the principles of quantum mechanics with those of general relativity has emerged as a central challenge in modern physics, inspiring extensive efforts to understand the influence of the structure of the spacetime on quantum phenomena. This quest has driven research into how its curvature, as described by various metric solutions, may subtly affect the behavior of quantum systems such as the energy levels of particles, wave functions, and even traditional interaction potentials themselves. 

In this context, the relativistic quantum mechanics in curved spacetimes \cite{Parker:1980hlc} has been extensively investigated in recent years, providing a large amount of results for different kinds of spacetime backgrounds. Some illustrative examples are static and rotating black hole metrics \cite{elizalde_1987, chandra}, the Hartle-Thorne spacetime \cite{Pinho:2023nfw} and also cylindrical symmetric metrics such as cosmic strings and other configurations \cite{Santos:2016omw,Santos:2017eef,Vitoria:2018its, Deglmann:2025mcl, Barbosa:2023gxl, Barbosa:2023rmq}. These investigations also encompass quantum oscillators \cite{Ahmed:2022tca,Ahmed:2023blw,Santos:2019izx,Yang:2021zxo,Soares:2021uep,Rouabhia:2023tcl}, the Casimir effect \cite{Santos:2018jba}, and other phenomena \cite{Sedaghatnia:2019xqb,Guvendi:2022uvz,Vitoria:2018mun,Barros:2004ta}, which have motivated numerous studies.

{\color{black}
In this context, relativistic quantum mechanics in curved spacetimes \cite{Barros:2004ta} has been extensively investigated, with foundational work on one-electron atoms in these environments \cite{Parker:1980hlc}. As illustrative examples, for static black hole metrics, exact series solutions for the Klein-Gordon equation were obtained in the Schwarzschild geometry \cite{elizalde_1987}, while rotating black holes were explored by solving the Dirac equation in the Kerr metric \cite{chandra}. Similarly, the behavior of spin-0 bosons near slowly rotating stars has been modeled using the Hartle-Thorne spacetime \cite{Pinho:2023nfw}.

Cylindrical symmetric metrics and topological defects are also equally rich grounds for research. For example, the dynamics of scalar fields have been analyzed in cosmic string spacetime under noninertial effects in \cite{Santos:2016omw}, with extensions in \cite{Santos:2017eef}, while rotating effects in spacetimes with space-like and spiral dislocations were studied in \cite{Vitoria:2018its}. Similar works were done recently, considering black strings within anisotropic quintessence \cite{Deglmann:2025mcl} and exact solutions in the Bonnor-Melvin universe \cite{Barbosa:2023gxl} with extension regarding the Klein-Gordon oscillator in \cite{Barbosa:2023rmq}. Such system has been widely adapted, for example, to investigate effects by non-trivial topologies in \cite{Ahmed:2022tca}, \cite{Ahmed:2023blw} and \cite{Santos:2019izx}, G\"odel-type universes in \cite{Yang:2021zxo}, Ellis-Bronnikov wormholes in \cite{Soares:2021uep}, and accelerating Rindler spacetimes in \cite{Rouabhia:2023tcl}. In general, these background geometries influence diverse phenomena, for example, recently, rotational effects on Casimir energy were explored in \cite{Santos:2018jba}, Aharonov-Bohm effect for bound states in \cite{Vitoria:2018mun}, and also fermion dynamics in both Som-Raychaudhuri \cite{Sedaghatnia:2019xqb} and defect-generated spacetimes \cite{Guvendi:2022uvz}.}

A further step in this kind of development is to consider the possibility of the influence of the metric on the interactions between particles. In \textcolor{black}{\cite{BunchParkerProp} and \cite{BunchProp2RenormPhi4:1981tr}} a representation for the Feynman propagator in the Riemann normal coordinates has been proposed, and based on this formalism, \textcolor{black}{as extensively explained in literature (such in the review \cite{Shapiro:2008sf-SemiclassicalGravReview} and in the books \cite{book:ParkerToms} and \cite{book:ShapiroGravity})}, in a recent work \cite{Zamperlini:2025nly} it was shown that the Yukawa potential acquires corrections if an arbitrary metric is considered. This offers an interesting perspective in the study of nuclear interactions of particles inside strong gravitational fields, suggesting a way for these particles to probe spacetime curvature quantum mechanically, affecting fundamental interactions and generating corrections to the potential, \textcolor{black}{similar works regarding the Newtonian potential can be seen in \cite{GravitonNewtonMassesBjerrum-Bohr:2002gqz} and \cite{BornApproximationDeSitterFerrero:2021lhd}}.

It is then natural to investigate how different systems with strong gravitational fields might influence the various particle interactions. For instance, considering the Yukawa potential, neutron stars \textcolor{black}{(reviewed in \cite{ETL-Ozel:2016oaf}, \cite{ETL-hamel:2008ca}, \cite{ETL-Lattimer:2000nx}, and \cite{Menezes:2021jmw})} provide a natural setting for studying how the metric affects nucleon interactions. This paper extends the investigation of curvature-induced modifications presented in \cite{Zamperlini:2025nly}
by examining them within the Tolman metric framework \cite{Tolman:1939jz}, a static, spherically symmetric perfect fluid solution often used as a first approximation for describing compact stars in General Relativity. As a first approach, the $\Phi\Phi$ interaction in a $\phi^3$-like theory is considered as far as it generates a Yukawa-like potential and allows the calculation of the corrections of the potential in terms of an analytical expression, similar to the one obtained if the nucleon-nucleon interaction is taken into account, without the need of introducing spinors in the formalism. So, in this work, the Feynman amplitude for this interaction will be considered for an arbitrary spacetime with spherical symmetry, and then the Tolman IV and Tolman VI solutions will be used. With this procedure, the corrections to the propagator and then to the potential can be estimated by considering different values for the parameters of the theory.

This paper is structured as follows: Section \ref{sec:yukawapotTolman} reexamines the corrected Yukawa potential, explicitly deriving the interaction potential for generic spherically symmetric metrics and specifically for the Tolman IV and VI metrics, followed by a qualitative discussion of their effects. In Section \ref{sec:NumericalResults} these results are applied to known astrophysical objects with specific mass and radius values, providing numerical estimates of the expected energy shift magnitudes. Finally, Section \ref{sec:conclusions} presents our conclusions and the discussion of the results.

Throughout the paper, we make use of natural units, where $\hbar=c=1$ and $G=m_P^{-2} \approx 6.709\times10^{-45}\ \mathrm{MeV}^{-2}$, rendering all physical quantities in powers of energy.

\section{Yukawa potential within the context of Tolman metrics}\label{sec:yukawapotTolman}

In this work, we are interested in studying the corrections for a Yukawa-type potential for particles interacting in a Tolman metric. We will consider the approach proposed in \textcolor{black}{\cite{BunchParkerProp} and \cite{BunchProp2RenormPhi4:1981tr}, and explored in \cite{Zamperlini:2025nly}}, where the system is described in terms of Riemann normal coordinates. By calculating the propagator in an arbitrary metric, the potential of such an interaction in the Born approximation may be determined. In this section, we will show these expressions and then calculate them for general static spherically symmetric spaces. These results will be applied to the Tolman IV and Tolman VI solutions.

The Green function for spin-0 particles $G_F(x,x')$ may be determined by the equation
\begin{equation}
    \l(\Box_c + m^2 \r)G_F(x,x') = -\frac{\delta^{(4)}\l(x-x'\r)}{\sqrt{-g}} \quad,
\end{equation}
where $\Box_c=g^{\mu\nu}\d_\mu\d_\nu$ is the curved space d'Alembertian and $m$ is the $\phi$ mass, and it may be expressed as
\begin{equation}
    G(y) = \int \dtfourk e^{i k \cdot y}G(k) = \int \dtfourk e^{i k \cdot y}\l[G_0(k) + G_1(k) + G_2(k) + \cdots \r] \quad,
\end{equation}
in terms of the flat-space propagator $G_0(k)$ and its corrections $G_j(k)$ for $j = 1,2,...$ relative to the curved space in the momentum representation. In a Riemann normal coordinate expansion, keeping up to first order in the Ricci tensor, it is given by
\begin{equation}
    G(k) = \frac{1}{k^2+m^2} + \frac{1}{3}\frac{\mathcal{R}'}{\l(k^2+m^2\r)^2} - \frac{2}{3}\frac{R'^{\mu\nu}k_\mu k_\nu}{\l(k^2+m^2\r)^3} + \mathcal{O}\l(k^{-5}\r) \quad,
    \label{eq:propEspCurvoMom}
\end{equation}
\textcolor{black}{as obtained in \cite{BunchParkerProp}, \cite{BunchProp2RenormPhi4:1981tr} and \cite{book:ShapiroGravity}}, 
where $R'^{\mu\nu}$ and $\rscalar'$ are respectively the Ricci tensor components and the Ricci scalar calculated at the spacetime point $x'$.

\begin{figure}[H]
    \centering
    \includegraphics[scale=1]{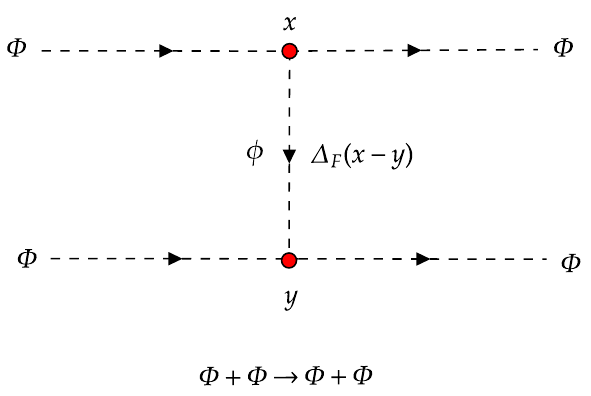}
    \caption{Leading order diagram for the process $\Phi+\Phi \ra \Phi+\Phi$.}
    \label{fig:diagramaphi3LO}
\end{figure}

Considering the $\phi^3$-like model with 
\begin{equation}
    \mathcal{L} = \frac{1}{2} (\partial_{\mu}\phi)^2 - \frac{\mu^2}{2} \phi^2 + \frac{1}{2} \partial_{\mu}\Phi^*\partial^{\mu}\Phi - \frac{m_{\Phi}^2}{2} |\Phi|^2 - \lambda \Phi^{*}\phi\Phi \qquad ,
    \label{eq:lagrangianphi3}
\end{equation}
and supposing the scattering $\Phi+\Phi \ra \Phi+\Phi$ at the tree level as it is shown in  the diagram of  \autoref{fig:diagramaphi3LO}, where $\phi$ represents the particle in the internal line, which can have a different mass, the Feynman amplitude is given by 
\begin{equation}
    \mathcal{M} = i \l(-i\lambda\r)^2\l(\frac{-i}{\mu^2-q^2}\r) \qquad ,
    \label{eq:amplitudephi3LO}
\end{equation}
\noindent
\textcolor{black}{in Minkowski space,} which determines the potential in the momentum space in a non-relativistic Born approximation
\begin{equation}
    \Tilde{V}\l(|\vq|\r) = \frac{\mathcal{M}}{4m_{\Phi}^2} \qquad,
    \label{eq:BornVq}
\end{equation}
\noindent
that is a Yukawa-like potential in the configuration space.

\textcolor{black}{Strictly speaking, in a general curved, non-homogeneous spacetime interior, constructing globally defined in/out free states is challenging. For this reason, we emphasize our assumption of a local inertial frame around the point $x'$ (where we build the Riemann normal coordinate patch from), within a normal neighborhood small compared to curvature scales, curvature and its gradients are slowly varying, so that one may define approximate local plane-wave states and employ the standard Born relation / Fourier transforms. In summary, this is a quasi-local approximation, and the results are valid only where these assumptions hold.}

\begin{figure}[H]
    \centering
    \includegraphics[width=0.6\linewidth]{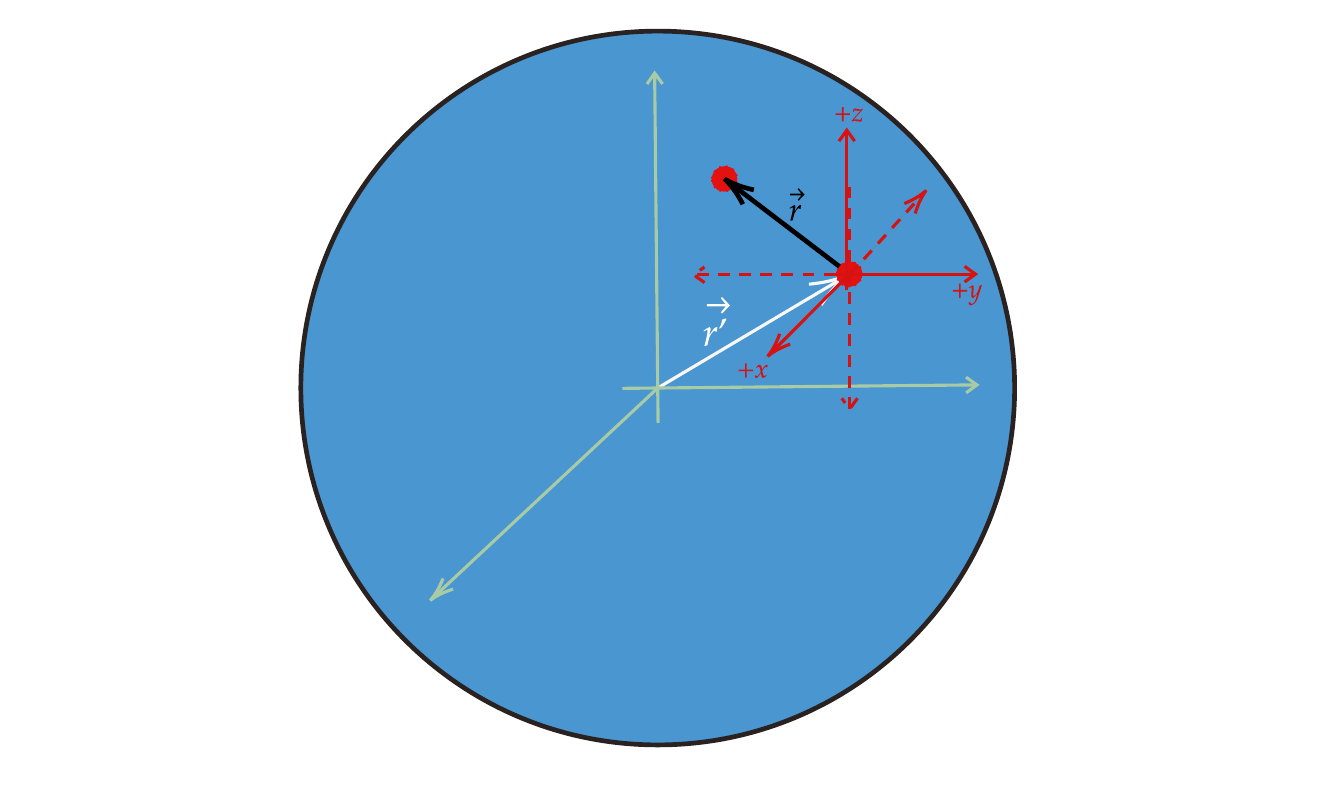}
    \caption{\textcolor{black}{Example of choice of coordinate system for two particles interacting via an interacting potential dependent on local position $\vec{r}$, in the interior of a spherically symmetric object of radius $r_\star$, around some point in the position $\vec{r'}$ relative to the metric origin.}}
    \label{fig:coordinatesdiagram}
\end{figure}

In a curved spacetime, considering \textcolor{black}{the local patch of Riemann coordinates,} from Eq. \ref{eq:propEspCurvoMom} we have
\cite{Zamperlini:2025nly}
\begin{equation}
    V(\vr) = -\frac{\lambda^2}{4m_\Phi^2}\int \dtq e^{i\vq\cdot\vr}\l[\frac{1}{\mu^2+|\vq|^2} + \frac{1}{3}\frac{\rscalar'}{\l(\mu^2+|\vq|^2\r)^2} - \frac{2}{3}\frac{R'_{\mu\nu}q^\mu q^\nu}{\l(\mu^2+|\vq|^2\r)^3} +\cdots \r] \quad,
    \label{PotentialCurvedSpaceInts}
\end{equation}
\noindent
where the first term generates a usual Yukawa-type potential and the other ones are given in terms of the corrections determined by $\rscalar'$ and ${R}'_{\mu\nu}$ \textcolor{black}{in the local inertial frame}. If we place a frame at a position $\vec{r'}$ from the origin, which represents the center of the stellar object, such as the one shown in \autoref{fig:coordinatesdiagram}, we may write the $\Phi\Phi$ potential with spacetime curvature corrections $V(g_{\mu\nu}',\vr)\equiv V(\vr)$ in terms of the flat-space potential $V_0(r)$:

{\color{black}
\begin{equation}
    \begin{aligned}
    V(\vr) = V_0(r) \vast\{  1 - \frac{1}{12} \biggr[&\l(-2 {\rscalar'}+{R'}_{(x)(x)} + {R'}_{(y)(y)} + {R'}_{(z)(z)}\r)\frac{r}{\mu} + \\ & -{R'}_{(x)(x)} x^2 - {R'}_{(y)(y)}y^2 - {R'}_{(z)(z)}z^2 +\\ & -2{R'}_{(x)(y)}xy - 2{R'}_{(x)(z)}xz - 2{R'}_{(y)(z)}yz\biggr] \vast\} \quad,
    \end{aligned}
\label{eq:potential-localcartesian}
\end{equation}
where the indices, or directional-axis $x,y,z$ are defined and interpreted relative to some choice of basis in the local frame. Alternatively, one can use the matrix notation for the position vector $\mathbf{x}$ and for the spatial sector of the Ricci tensor in these local cartesian coordinates $R_{ij}\equiv\mathbf{R}$, to rewrite \autoref{eq:potential-localcartesian} in compact form:
\begin{equation}
    \begin{aligned}
    V(\vr) = V_0(r) \left\{  1 - \frac{1}{12} \biggr[\l(-2 {\rscalar'}+\operatorname{tr}\mathbf{R}'\r)\frac{r}{\mu} -\mathbf x^{T}\mathbf{R}'\mathbf x \biggr] \right\} \quad.
    \end{aligned}
\label{eq:potential-localcartesian-compact}
\end{equation}
}

{\color{black}

\subsection{Yukawa potential in a general static spherically symmetric spacetime}

If one wishes to express the same potential, but within a spherically symmetric metric given by:
\begin{equation}
    \mathrm{d}s^2 = -e^{f(r)}\mathrm{d}t^2 + e^{g(r)}\mathrm{d}r^2 + r^2\mathrm{d}\theta^2 + r^2 \sin^2(\theta) \mathrm{d}\varphi^2 \quad,
    \label{eq:genericsphericalmetriclineelement}
\end{equation}
one can conveniently define these directions accordingly to the \autoref{fig:diagramaesfricovetorbase}, where the $y$-axis defines the axis connecting the Yukawa potential center and the metric center, or origin, defining a parallel (globally-radial) orientation $\hat{y}=\hat{r'}$; and then defining the $\hat{x}=\hat{\varphi'}$ and $\hat{z}=\hat{\theta'}$ directions as transversal ones.

\begin{figure}[H]
    \centering
    \includegraphics[width=\linewidth]{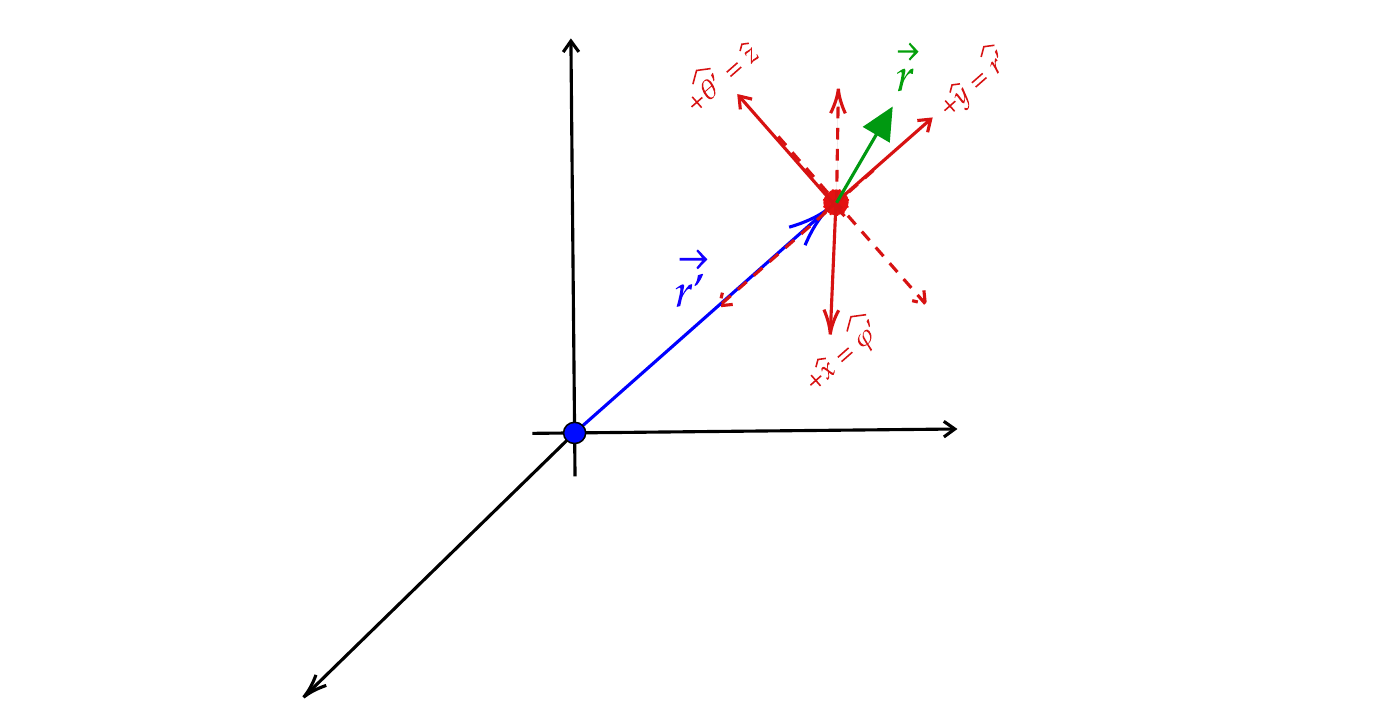}
    \caption{\textcolor{black}{Diagram of vectors and directions given by a spherical metric at point $x'$, depicting our choice for the local unit vectors orientations and labeling.}}
    \label{fig:diagramaesfricovetorbase}
\end{figure}

With this identification, one can rewrite \autoref{eq:potential-localcartesian-compact}, with respect to the labels from the spherical metric, as:
\begin{equation}
    V(\vec{r}) = V_0(r) \l\{1  - \frac{1}{12}\l[\l(-2 \rscalar' +R'_{(r)(r)} + R'_{(\theta)(\theta)} + R'_{(\varphi)(\varphi)}\r)\frac{ r}{\mu} - R'_{(r)(r)} y^2  - R'_{(\theta)(\theta)}z^2 -R'_{(\varphi)(\varphi)}x^2 \r]\r\} \quad,
    \label{eq:YukawaCurvedGeneric}
\end{equation}
where the local spherical components of the Ricci tensor, $R'_{(i)(j)}$, are related to the global coordinate components, $R_{\mu\nu}$, via the projection onto the local inertial frame through
\begin{equation}
    R_{(a)(b)}=e_{(a)}{}^\mu e_{(b)}{}^\nu R_{\mu\nu} \quad,
    \label{eq:RicciLocalBasis}
\end{equation}
via the spatial tetrad (\textit{dreibein}):
\begin{equation}
\begin{aligned}
    e_{(r)}{}^\mu(x') &= \l(\,0,\,e^{-g(r')/2},\,0,\,0\,\r)\quad;\\
    e_{(\theta)}{}^\mu(x') &= \l(\,0,\,0,\,1/r',\,0\,\r)\quad;\\
    e_{(\varphi)}{}^\mu(x') &= \l(\,0,\,0,\,0,\,1/\left(r'\sin\theta'\right)\,\r) \quad,
    \label{eq:tetradaesfericaespacial}
\end{aligned}
\end{equation}
such that the spatial coordinate components for the metric given in \autoref{eq:genericsphericalmetriclineelement} are then:
\begin{align}
    R_{rr}&= - \frac{\left(\frac{d}{d r} f{\left(r \right)}\right)^{2}}{4} + \frac{\frac{d}{d r} f{\left(r \right)} \frac{d}{d r} g{\left(r \right)}}{4} - \frac{\frac{d^{2}}{d r^{2}} f{\left(r \right)}}{2} + \frac{\frac{d}{d r} g{\left(r \right)}}{r} \label{eq:Rrrcomponent}\\
    R_{\theta \theta}&= \frac{\left(- r \frac{d}{d r} f{\left(r \right)} + r \frac{d}{d r} g{\left(r \right)} + 2 e^{g{\left(r \right)}} - 2\right) e^{- g{\left(r \right)}}}{2} \label{eq:Rthetathetacomponent}\\
    R_{\varphi \varphi}&= \frac{\left(- r \frac{d}{d r} f{\left(r \right)} + r \frac{d}{d r} g{\left(r \right)} + 2 e^{g{\left(r \right)}} - 2\right) e^{- g{\left(r \right)}} \sin^{2}{\left(\theta \right)}}{2} = \sin^2(\theta)\, R_{\theta\theta} \label{eq:Rvarphivarphicomponent}
\end{align}
and the corresponding curvature scalar is
\begin{equation}
    \mathcal{R}=\frac{\left(- r^{2} \left(\frac{d}{d r} f{\left(r \right)}\right)^{2} + r^{2} \frac{d}{d r} f{\left(r \right)} \frac{d}{d r} g{\left(r \right)} - 2 r^{2} \frac{d^{2}}{d r^{2}} f{\left(r \right)} - 4 r \frac{d}{d r} f{\left(r \right)} + 4 r \frac{d}{d r} g{\left(r \right)} + 4 e^{g{\left(r \right)}} - 4\right) e^{- g{(r )}}}{2 r^{2}} \quad.
\end{equation}

One can show, from Eqs. \ref{eq:tetradaesfericaespacial}, \ref{eq:Rrrcomponent}, \ref{eq:Rthetathetacomponent} and \ref{eq:Rvarphivarphicomponent}, that \autoref{eq:RicciLocalBasis} will yield:
\begin{equation}
\begin{aligned}
    R'_{(r)(r)} &= e^{-g(r)}{R'}_{rr}\quad; \\
    R'_{(\theta)(\theta)}& = \frac{{R'}_{\theta\theta}}{r'^2}\quad;\\
    R'_{(\varphi)(\varphi)} &= \frac{R_{\varphi\varphi}}{r'^2\sin^2\theta'}=\frac{R'_{\theta\theta}(r')\sin^2\theta'}{r'^2\sin^2\theta'}=R'_{(\theta)(\theta)} \quad.
\end{aligned}
\label{eq:RicciLocalSpherical}
\end{equation}

\autoref{eq:RicciLocalSpherical} explicitly shows the \emph{normalization} by metric factors between coordinate components and ``measured'' components in the inertial frame.

Given the configuration of the local curvature \ref{eq:RicciLocalSpherical}, we can define
\begin{equation}
R_{(r')(r')} = R_{\parallel}\quad,\quad R_{(\theta')(\theta')}=R_{(\varphi')(\varphi')}=R_{\perp} \quad,
\end{equation}
whose spatial trace \( \tr\mathbf{R} = R_{(r)(r)}+R_{(\theta)(\theta)}+R_{(\varphi)(\varphi)}=R_{\parallel}+2R_{\perp}\).

This allows to rewrite \autoref{eq:YukawaCurvedGeneric} as \footnote{Using that $x^2+z^2=r^2-y^2$.}:
\begin{equation}
    V(\vec{r}) = V_0(r) \l\{1  + \frac{1}{12}\l[\l(2 \rscalar' - \tr\mathbf{R}\r)\frac{ r}{\mu} + R_{\perp}r^2 + (R_{\parallel} - R_{\perp})y^2\r]\r\} \quad,
    \label{eq:YukawaCurvedGeneric-parallelperp}
\end{equation}
which shows that if the local curvature $R_{\parallel}\neq R_{\perp}$, the original potential loses its radial symmetry due to the spacetime structure. And if $R_{\parallel} = R_{\perp}$, it becomes:
\begin{equation}
    V(\vec{r}) = V_0(r) \l\{1  + \frac{1}{12}\l[\l(2 R' - 3 R_{\parallel} \r)\frac{ r}{\mu} + R_{\parallel}r^2 \r]\r\} \quad.
    \label{eq:potential-Isotropic}
\end{equation}
 
}

From Eq. \ref{eq:YukawaCurvedGeneric-parallelperp}, to have significant curvature effects, the relevant Ricci tensor components and curvature scalar must be on the order of the inverse square of the coordinates probed by the interacting particle. For nuclear interactions, such corrections become relevant when the spacetime curvature quantities are approximately $10\ \mathrm{fm}^{-2}$ ($\sim (20\ \mathrm{MeV})^2$ in natural units).

Focusing on the spacetime curvature corrections for the Yukawa potential, it is interesting to define
\begin{equation}
    \frac{\Delta V}{V_0} = \frac{V(g_{\mu\nu}, \vr) - V_0(r)}{V_0(r)} \quad,
\end{equation}
as a quantity to study the magnitude of the corrections to the potential, which will be given by: 
{\color{black}
\begin{equation}
    \frac{\Delta V}{V_0} = \frac{1}{12}\l[\l(2 \rscalar' - \tr\mathbf{R}\r)\frac{ r}{\mu} + R_{\perp}r^2 + (R_{\parallel} - R_{\perp})y^2\r] \quad.
    \label{eq:potesfericasSPECIFICPOSITION}
\end{equation}
}

Eq. \ref{eq:potesfericasSPECIFICPOSITION} can be used to study the curvature corrections of the Yukawa potential for any spherically symmetric metric by replacing the functions $f(r)$ and $g(r)$. \textcolor{black}{We interpret $\Delta V$ as the energy shift experienced by a commoving (or local) observer in the immediate normal neighborhood of the interacting particles (i.e. in the local inertial patch with origin in the potential source $x'$). This is the frame in which the potential is defined locally.}

Now we focus our attention on two specific Tolman solutions, which are used as models of compact stars in many studies, by applying the expressions to known astrophysical objects.

\subsection{Tolman-IV solution}

One of the most reasonable Tolman solutions is the fourth one presented in his paper, which displays a sensible solution of a compressible fluid sphere with the pressure dropping to zero at the surface, with appropriate boundary conditions. This solution serves as a good approximation to describe compact stars, as shown recently in \cite{TolmanIVApp_Panotopoulos:2024imo}.

The assumption for the Tolman IV solution is that $e^{f} \frac{f'}{2r} = \text{constant}$, which for the metric implies that 
\begin{equation}
    e^{g(r)} = \frac{1 + 2r^2/A^2}{(1-r^2/R^2)(1+r^2/A^2)} \qquad\text{and}\qquad e^{f(r)}=B^2(1+r^2/A^2) \quad,
\end{equation}
that provides the specific forms for the pressure and energy density given by:
\begin{equation}
\begin{aligned}
    8 \pi Gp &= \frac{1}{A^2}\l(\frac{1-A^2/R^2-3r^2/R^2}{1+2r^2/A^2}\r)+\Lambda \quad,\\
    8 \pi G\rho &= \frac{1} {A^2}\frac{1+3A^2/R^2+3r^2/R^2}{1+2r^2/A^2}+\frac{2}{A^2}\frac{1-r^2/R^2}{(1+2r^2/A^2)^2} - \Lambda \quad,
\end{aligned}
\end{equation}
with $A$, $R$, and $B$ being constants to be found or fitted to data.

By constraining the pressure to be null at the boundary at the star surface $r=r_\star$, we find that
\begin{equation}
    r_\star = \frac{R}{\sqrt{3}} \sqrt{1-\frac{A^2}{R^2}} \quad,
    \label{eq:radiusTolmanIV}
\end{equation}
and by setting the metric to connect the exterior Schwarzschild solution, the mass of the sphere must be
\begin{equation}
    M = \frac{r_\star}{2G}\l[1-\frac{\l(1-r_\star^2/R^2\r)\l(1+r_\star^2/A^2\r)}{1+2r_\star^2/A^2}\r] \quad.
    \label{eq:massTolmanIV} 
\end{equation}

Within this metric framework, one can find the relation between $A$ and $R$ from Eq. \ref{eq:radiusTolmanIV}:
\begin{equation}
    A = \sqrt{R^2-3r_\star^2} \quad,
    \label{eq:AdeR-TolmanIV}
\end{equation}
and by substituting Eq. \ref{eq:AdeR-TolmanIV} in  \ref{eq:massTolmanIV}, one can find the relationship between mass and radius depending on $R$, which can be calculated from
{\color{black}
\begin{equation}
    R^2 = \frac{r_\star^3}{GM} = \frac{2\,r_\star^3}{r_s} \quad,
    \label{eq:Rdependederstar-TolmanIV}
\end{equation}
where $r_s=2GM$ is the Schwarzschild radius for the mass $M$.}

So, by having the mass and radius as input one can find the values of the parameters $A$ and $R$, which are the ones that are relevant for this work \footnote{The last parameter $B$, can also be found by connecting with Schwarzschild solution, and will be given by $B=\sqrt{\l(1-r_\star^2/R^2\r)\l(1+2r_\star^2/A^2\r)}$.}.

Calculating the \textcolor{black}{coordinate} components of the Ricci tensor for this specific solution, considering $\Lambda=0$, we have
\begin{equation}
\begin{aligned}
    R'_{rr} &= \frac{ \l(- 2 A^{4} - A^{2} R^{2} - 6 A^{2} r'^{2} - 6 r'^{4}\r)}{ \l(- A^{4} R^{2} + A^{4} r'^{2} - 3 A^{2} R^{2} r'^{2} + 3 A^{2} r'^{4} - 2 R^{2} r'^{4} + 2 r'^{6}\r)}  \quad; \\
    \rscalar' &= \frac{2 \l(3 A^{4} + 11 A^{2} r'^{2} - 2 R^{2} r'^{2} + 12 r'^{4}\r)}{R^{2} \l(A^{4} + 4 A^{2} r'^{2} + 4 r'^{4}\r)} \quad; \\
    \frac{R'_{\theta\theta }}{r'^2} &= \frac{ \l(2 A^{4} + A^{2} R^{2} + 6 A^{2} r'^{2} + 6 r'^{4}\r)}{ R^{2} \l(A^{4} + 4 A^{2} r'^{2} + 4 r'^{4}\r)} \quad, 
\end{aligned}
\label{eq:correctionsTolmanIV}
\end{equation}
{\color{black}
and in the local basis at the expansion point at $r'$ we will have
\begin{equation}
    R_\parallel = R_\perp = \frac{2 A^{4} + A^{2} R^{2} + 6 A^{2} r'^{2} + 6 r'^{4}}{R^{2} \l(A^{2} + 2 r'^{2}\r)^{2}} \quad,
\label{eq:correctionsTolmanIV-localbasis}
\end{equation}
such that the interacting potential at \autoref{eq:YukawaCurvedGeneric-parallelperp} becomes
\begin{equation}
\begin{aligned}
    V(r) = V_0(r) \vast\{&  1 +  \frac{1}{12} \biggr[{2} {\frac{2 \l(3 A^{4} + 11 A^{2} r'^{2} - 2 R^{2} r'^{2} + 12 r'^{4}\r)}{R^{2} \l(A^{2} + 2 r'^{2}\r)^{2}}} +\\& -  \frac{2 A^{4} + A^{2} R^{2} + 6 A^{2} r'^{2} + 6 r'^{4}}{R^{2} \l(A^{2} + 2 r'^{2}\r)^{2}} (3-\mu r)\biggr]\frac{r}{\mu} \vast\} \quad,
\end{aligned}
\label{eq:potential-TolmanIV-v1}
\end{equation}
or separating into terms proportional to powers of $r$ inside the brackets:
\begin{equation}
\begin{aligned}
    V(r) = V_0(r) \Biggr\{  1 +  \frac{1}{12} \biggr[&\frac{6 A^4 + 26 A^2{r'}^2 - 3A^2R^2-8 R^2 {r'}^2 + 30{r'}^4}{R^{2} \l(A^{2} + 2 r'^{2}\r)^{2}}\frac{r}{\mu}  +\\& +  \frac{2 A^{4} + A^{2} R^{2} + 6 A^{2} r'^{2} + 6 r'^{4}}{R^{2} \l(A^{2} + 2 r'^{2}\r)^{2}} {r^2}\biggr]\Biggr\} \quad.
\end{aligned}
\label{eq:potential-TolmanIV}
\end{equation}

So, the curvature corrections still leave the interaction with radial symmetry, which must be related to the isotropic nature of the matter content that provides the Tolman metrics, coded in the isotropic pressure in the energy-momentum tensor of the spherical fluid \cite{Tolman:1939jz}.
}

\subsubsection{Limit $r'\ra0$}

We can expect the largest corrections to be at the center of the spherical object, where the pressure is highest. The pressure and density at the center are given by (with $\Lambda=0$)
\begin{align}
    p_c &= \frac{1-A^2/R^2}{8 \pi GA^2} \quad,\\
    \rho_c &= \frac{3+3A^2/R^2}{8 \pi GA^2} \quad.
\end{align} 

{\color{black}
From the corrections shown in Equations \ref{eq:correctionsTolmanIV} and \ref{eq:correctionsTolmanIV-localbasis}, taking the limit $r'\ra 0$, we have

\begin{align}
    \lim_{r'\ra 0} \rscalar' &= \frac{6}{R^{2}} \quad;\\
    \lim_{r'\ra 0} R_{\parallel} &= \frac{2}{R^{2}} + \frac{1}{A^{2}}  \quad.
\end{align}

Thus, if the interacting system is located at the center of the spherical object, the corrected Yukawa potential reads as:

\begin{equation}
    V(r) = V_0(r) \l\{1  + \frac{1}{12}\l[\l(\frac{6}{R^{2}} - \frac{3}{A^{2}}\r) \frac{r}{\mu} + \l(\frac{2}{R^{2}} + \frac{1}{A^{2}}\r) r^2 \r]\r\} \quad,  
    \label{eq:potencialrprime0}
\end{equation}

To simplify, we define:
\begin{equation}
    \begin{aligned}
    R_1 &= \frac{6}{R^{2}} - \frac{3}{A^{2}} \quad;\\ 
    R_2 &= \frac{2}{R^{2}} + \frac{1}{A^{2}} \quad.
    \end{aligned}
    \label{eq:CorrectionR1R2-TolmanIV}
\end{equation}
}

{\color{black}
From here onward, we rewrite the expressions in \eqref{eq:CorrectionR1R2-TolmanIV} explicitly in terms of \(r_\star\) and the corresponding \(r_s\), from the mass. Using \eqref{eq:AdeR-TolmanIV} and \eqref{eq:Rdependederstar-TolmanIV}, we obtain after simplifications:
\begin{equation}
    \begin{aligned}
    R_1 &= \frac{6}{R^{2}} - \frac{3}{A^{2}}
         \;=\; \frac{3\,r_s\,(r_\star - 3r_s)}{r_\star^{3}\,\bigl(2r_\star - 3r_s\bigr)} \;, \\[6pt]
    R_2 &= \frac{2}{R^{2}} + \frac{1}{A^{2}}
         \;=\; \frac{3\,r_s\,(r_\star - r_s)}{r_\star^{3}\,\bigl(2r_\star - 3r_s\bigr)} \;.
    \end{aligned}
    \label{eq:R1R2-rstar-rs-TolmanIV}
\end{equation}

These final forms highlight a common factor \(\tfrac{3r_s}{r_\star^{3}\,(2r_\star - 3r_s)}\), facilitating the discussion about the signs of \(R_1\) and \(R_2\). In particular, it is observed that \(R_1\) vanishes for \(r_\star = 3r_s\) and changes sign for larger radius values, while \(R_2\) vanishes at \(r_\star = r_s\). The condition \(2r_\star - 3r_s = 0\) characterizes a mathematical singularity of the model (\(r_\star = \tfrac{3}{2}r_s\)), causing modeled objects with a radius between $1.5$ and $3\,r_s$ to yield $R_1<0$ and $R_2>0$, while above $3\ r_s$ both corrections are positive, as we can infer from \autoref{fig:r1r2_rstar_rs_tolmanIV}.

\begin{figure}[H]
    \centering
    \includegraphics[width=\linewidth]{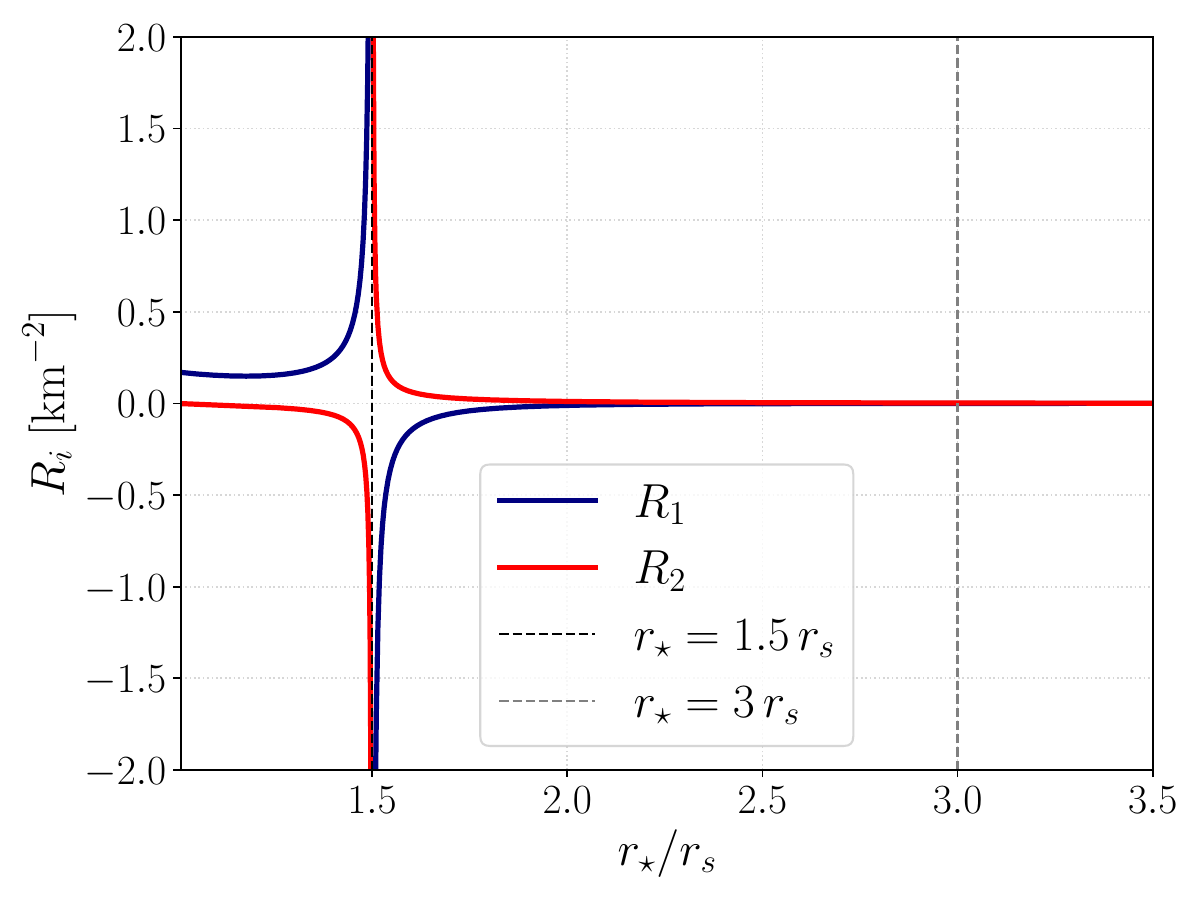}
    \caption{\textcolor{black}{Values of $R_1$ and $R_2$ depend on $r_\star/r_s$, that is, the compactness. For this example, a model of $M=2\,M_\odot$ was considered.}}
    \label{fig:r1r2_rstar_rs_tolmanIV}
\end{figure}

Thus, we rewrite the Yukawa potential at the highest pressure point in Tolman-IV:
\begin{equation}
    V(r) = V_0(r) \l\{1  + \frac{1}{12}\frac{3\,r_s}{r_\star^{3}\,\bigl(2r_\star - 3r_s\bigr)}\l[(r_\star - 3r_s)\frac{r}{\mu} +  (r_\star - r_s)r^2 \r]\r\} \quad,
    \label{eq:potencialTolmanIVrprime0Simplificado}
\end{equation}
or
\begin{equation}
    V(r) = V_0(r) \l\{1  + \frac{1}{12}\frac{3\,r_s}{r_\star^{3}\,\bigl(2r_\star - 3r_s\bigr)}\l[r_\star\l(\frac{r}{\mu}+r^2\r) - r_s\l(3\frac{r}{\mu}+r^2\r)  \r]\r\} \quad.
    \label{eq:potencialTolmanIVrprime0SimplificadoAlternative}
\end{equation}

Thus, the practical effect of the curvature on the interaction depends on the particle's position $r$, but mainly on the compactness of the modeled object, encoded in $r_\star/r_s$. 
}

\subsection{Tolman VI}

To compare different Tolman solutions, we analyze the Tolman-VI one, which provides a simpler expression for the dependence of $p$ on $r$, making it an ideal approach to investigate fluid spheres with infinite density and pressure at the center, as Tolman further discusses in his paper.

The assumption for the solution VI is that $e^{-g} = \text{constant}$, which for the metric implies that 
\begin{equation}
    e^{f} = \l(A r^{1-n}-Br^{1+n}\r)^2 \qquad\text{and}\qquad e^{g} = 2-n^2 \quad,
\end{equation}
providing a specific form for the pressure and energy density given by
\begin{equation}
\begin{aligned}
    8 \pi Gp &= \frac{1}{2-n^2}\frac{1}{r^2}\frac{(1-n)^2A-(1+n)^2Br^{2n}}{A-Br^{2n}}+\Lambda \quad,\\
    8 \pi G\rho &= \frac{1-n^2}{2-n^2}\frac{1}{r^2}-\Lambda \quad,
\end{aligned}
\end{equation}
with $n$, $A$, and $B$ being constants to be found, adjusted, or fitted to data. And again, for simplicity reasons, we will take $\Lambda=0$ here onward. 

This solution also differs due to a singularity at the center, which can be inferred from the corresponding Kretschmann scalar:
{\color{black}
\begin{equation}
    K(n=1/2) = \frac{24 \left(3 A^{2} - 10 A B r + 15 B^{2} r^{2}\right)}{49 r^{4} \left(A^{2} - 2 A B r + B^{2} r^{2}\right)} \quad.
\end{equation}}

As Tolman himself argues in Section 8 of his paper, a sensible choice is $n=1/2$, which results in the relationship
\begin{equation}
    \frac{p_c}{\rho_c}=\frac{1}{3} \quad,
\end{equation}
with $p_c$ and $\rho_c$ being the pressure and energy density at the center ($r=0$), respectively.

With this choice, the pressure and energy density are:
\begin{equation}
\begin{aligned}
    8 \pi Gp &= \frac{1}{7r^2}\frac{A-9Br}{A-Br} \quad,\\
    8 \pi G\rho &= \frac{3}{7r^2} \quad,
\end{aligned}
\end{equation}
and the boundary of the fluid sphere is given by
\begin{equation}
    r_\star=\frac{A}{9B} \quad,
\end{equation}
where $A$ has dimensions $L^{-1/2}$ and $B$ has dimensions $L^{-3/2}$.

As Tolman analyzed, this metric fixes the mass-radius ratio as 
\begin{equation}
    \frac{M}{r_\star}=\frac{3}{14G} \ ,
    \label{eq:massradiustolman6}
\end{equation}
and
from the radius relationship, we know how to obtain $A$,
\begin{equation}
    A= 9Br_\star \quad.
    \label{eq:AdeTolman6}
\end{equation}

By matching the Schwarzschild solution at the boundary, from the time component of the metric $e^{f(r)}$, and substituting $A$, we have
\begin{equation}
    \l(9Br_\star \cdot r_\star^{1/2} - Br_\star^{3/2}\r)^2 = 1 - \frac{2GM}{r_\star} \quad,
\end{equation}
and simplifying this expression using Eq.\ref{eq:massradiustolman6} results in:
\begin{equation}
    \l(8Br_\star^{3/2}\r)^2 = \frac{4}{7} \quad,
\end{equation}
which implies:
\begin{equation}
    B = \frac{1}{\sqrt{112}r_\star^{3/2}}=\frac{\sqrt{7}r_\star^{-3/2}}{28} \quad.
    \label{eq:BdeTolman6}
\end{equation}

Within this choice of $n$, the metric components become
\begin{equation}
    e^{f} = \l(A r^{1/2}-Br^{3/2}\r)^2 \qquad\text{and} \quad e^{g} = 7/4 \quad,
\end{equation}
such that the curvature quantities become:
{\color{black}
\begin{align}
    \rscalar' &= \frac{24 B}{7 {r}' \left(A-Br'\right)} \quad;\\
    R'_{rr} &= \frac{ A + 3 B {r}'}{4 {r}'^{2} \left(A - B {r}'\right)} \quad;\\    
    R'_{\theta\theta } &= \frac{A + 3 B {r}'}{7 \left(A - B {r}'\right)} \quad,
\end{align}
and in the local base we have
\begin{equation}
    R_\parallel=R_\perp = \frac{A + 3 B {r'}}{7 {r'}^{2} \left(A - B {r'}\right)}
\end{equation}
and then the curvature-corrected Yukawa potential on \autoref{eq:YukawaCurvedGeneric-parallelperp} will be 
\begin{equation}
    V(r) = V_0(r) \l\{  1 +  \frac{1}{12} \biggr[{\frac{48 B}{7 r' \left(A-Br'\right)} }  
    -  \frac{A + 3 B {r'}}{7 {r'}^{2} \left(A - B {r'}\right)} (3-\mu r)\biggr]\frac{r}{\mu} \r\} \quad,
\label{eq:potential-TolmanVI-v1}
\end{equation}
or, separating into terms proportional to powers of $r$ inside the brackets, like \autoref{eq:potential-Isotropic}:
\begin{equation}
\begin{aligned}
    V(r) = V_0(r) \l\{  1 +  \frac{1}{12} \biggr[{\frac{39 B {r'} - 3A}{7 {r'}^2 \left(A-Br'\right)} }\frac{r}{\mu}  
    + \frac{A + 3 B {r'}}{7 {r'}^{2} \left(A - B {r'}\right)} r^2\biggr] \r\} \quad.
\end{aligned}
\label{eq:potential-TolmanVI}
\end{equation}

Using Eqs. \ref{eq:AdeTolman6} e \ref{eq:BdeTolman6}, we can rewrite \autoref{eq:potential-TolmanVI} in terms only of the radius of the fluid sphere $r_\star$:
\begin{equation}
\begin{aligned}
    V(r) = V_0(r) \l\{  1 +  \frac{1}{12} \l[{\frac{39  \frac{r'}{r_\star} - 27}{7 {r'}^2 \left(9-\frac{r'}{r_\star}\right)} }\frac{r}{\mu}  
    + \frac{9+3\frac{r'}{r_\star}}{7 {r'}^{2} \left(9-\frac{r'}{r_\star}\right)} r^2\r] \r\} \quad,
\end{aligned}
\label{eq:potential-TolmanVI-rstar}
\end{equation}
from which we see that the correction linear in local radial distance $r$ is positive for $r'/r_\star>27/39$ and negative otherwise, indicating an attractive/repulsive effect dependent on the local frame position $r'$ for this order; whereas the correction quadratic in local radial distance is always positive, since $r'/r_\star\leq 1$, providing an always attractive feature for this order. The full effect is going to be a distortion of the potential dependent on the local radial distance and the local frame position, being analyzed in the next section.
}

At this point, it is interesting to investigate the corrections calculated for the potential in Tolman IV and in Tolman VI solutions. For this purpose, a numerical analysis for reasonable values of the parameters must be carried out, and it will be done in the next section.

\newpage
\section{Numerical Results}\label{sec:NumericalResults}

\subsection{Tolman-IV Numerical Results}

In this section, we will explore the dependence of the corrections to the potential on the parameters and on the distance from the origin for the considered Tolman solutions. The parameters that describe observed astrophysical objects will be considered. 

{\color{black}
Quantitatively, we can analyze the magnitude of the corrections in terms of the variation of $r'$. Following the results of \cite{TolmanIVApp_Panotopoulos:2024imo}, we plot the curvature corrections for the linear and quadratic order in $r$, seen in \autoref{eq:potential-Isotropic}, for two objects fitted to the Tolman-IV metric: Massive Pulsar J0740+6620 (\autoref{fig:curvaturerprimePulsar}) and the Compact Object HESS J1731-347 (\autoref{fig:curvaturerprimeHESS}), which show values on the order of $10^{-34}$--$10^{-32}$ $\mathrm{MeV^2}$ for most of the objects' radii.

\begin{figure}[H]
    \centering
    \includegraphics[width=\linewidth]{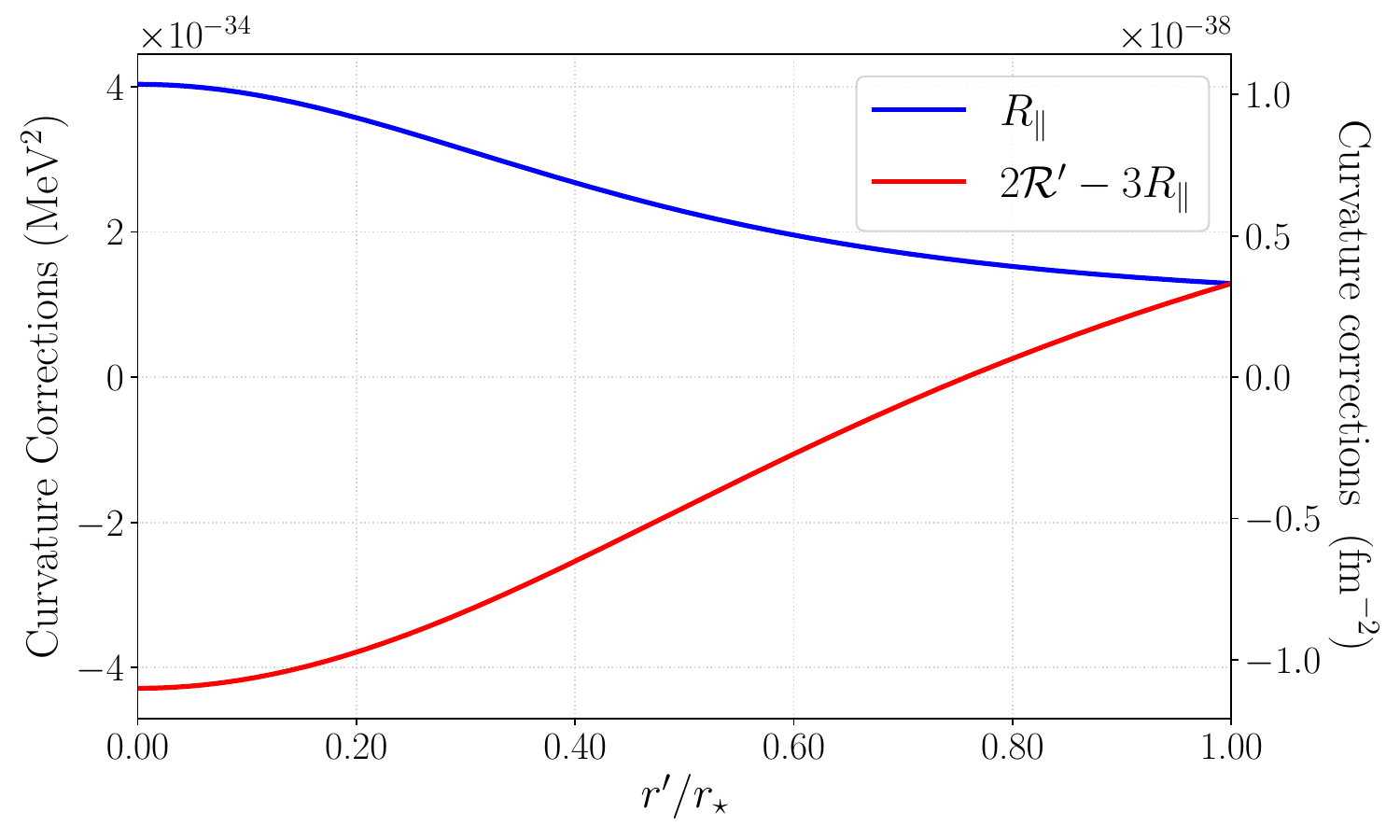}
    \caption{\textcolor{black}{Values of the magnitude of the curvature corrections for the Yukawa potential for a range of values of $r'/r_\star$ for the object Pulsar J0740+6620 with mass $M=2.1 M_\odot$ and a radius $r_\star=12.32\ \mathrm{km}$.}}
    \label{fig:curvaturerprimePulsar}
\end{figure}

\begin{figure}[H]
    \centering
    \includegraphics[width=\linewidth]{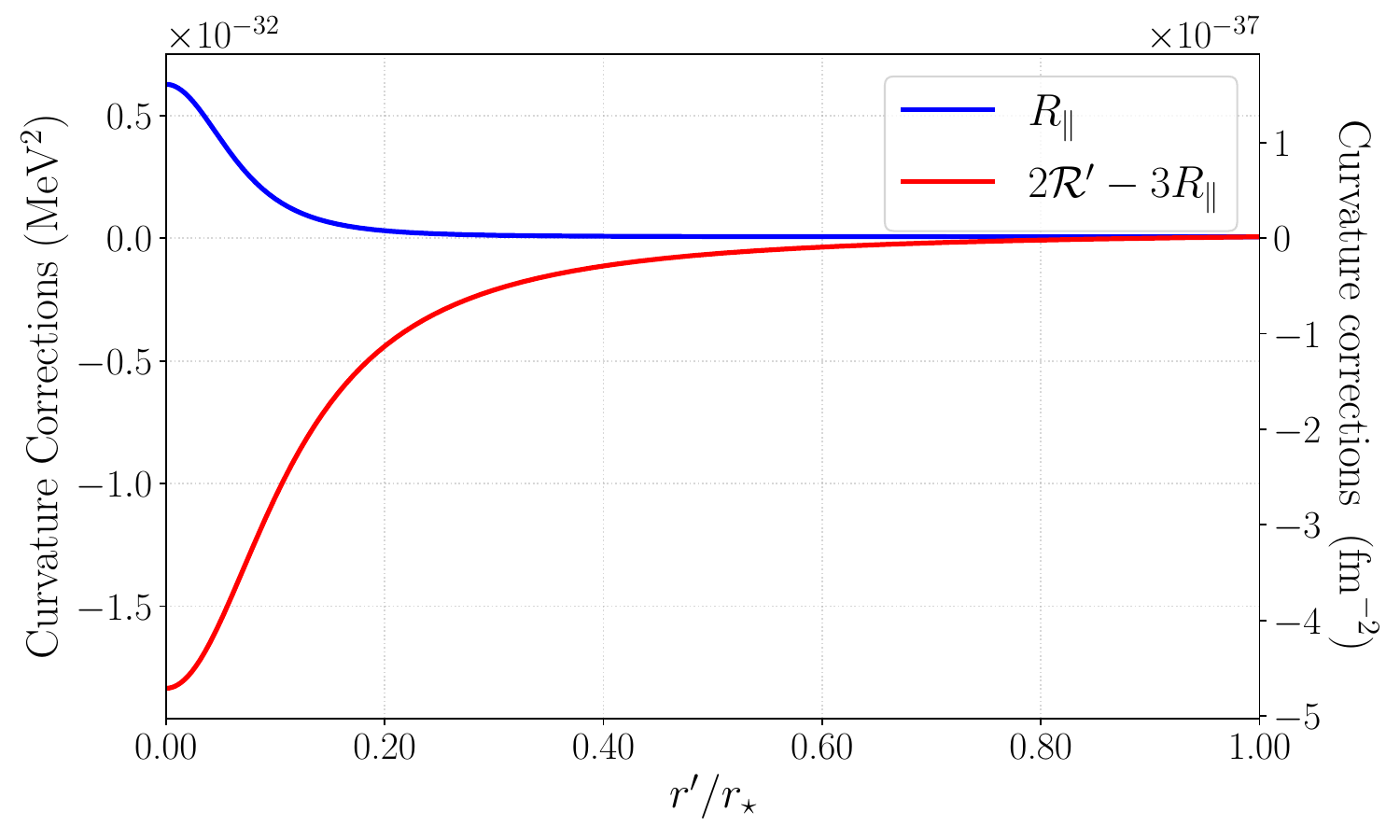}
    \caption{\textcolor{black}{Values of the magnitude of the curvature corrections for the Yukawa potential for a range of values of $r'/r_\star$ for the Compact Object HESS J1731-347 with mass $M=0.8\ M_\odot$ and a radius $r_\star=10.42\ \mathrm{km}$.}}
    \label{fig:curvaturerprimeHESS}
\end{figure}
}

\textcolor{black}{Moreover, the curvature magnitudes plotted in Figs. \ref{fig:curvaturerprimePulsar} and \ref{fig:curvaturerprimeHESS} are many orders of magnitude below the energy scale probed by the Yukawa exchange in our analysis (Fermi scale). Consequently, curvature gradients are negligible across the stellar parameter range shown, and the Riemann-normal-coordinate local expansion we employ to extract the effective potential is well justified within the small patches probed.}

{\color{black} The curvature corrections at the center of the compact object itself, as predicted by \autoref{eq:potencialrprime0} and defined on \autoref{eq:CorrectionR1R2-TolmanIV}, are expected to become significant at the nuclear interaction scale (around $10^{-15}\ \mathrm{m} = 10^{-18} \ \mathrm{km}$). \autoref{fig:curvatures3DReA} confirms this prediction, showing corrections of the order of $\mathrm{MeV^2}$ for $A$ and $R$ near the Fermi scale.
\begin{figure}[H]
    \centering
    \includegraphics[width=\linewidth]{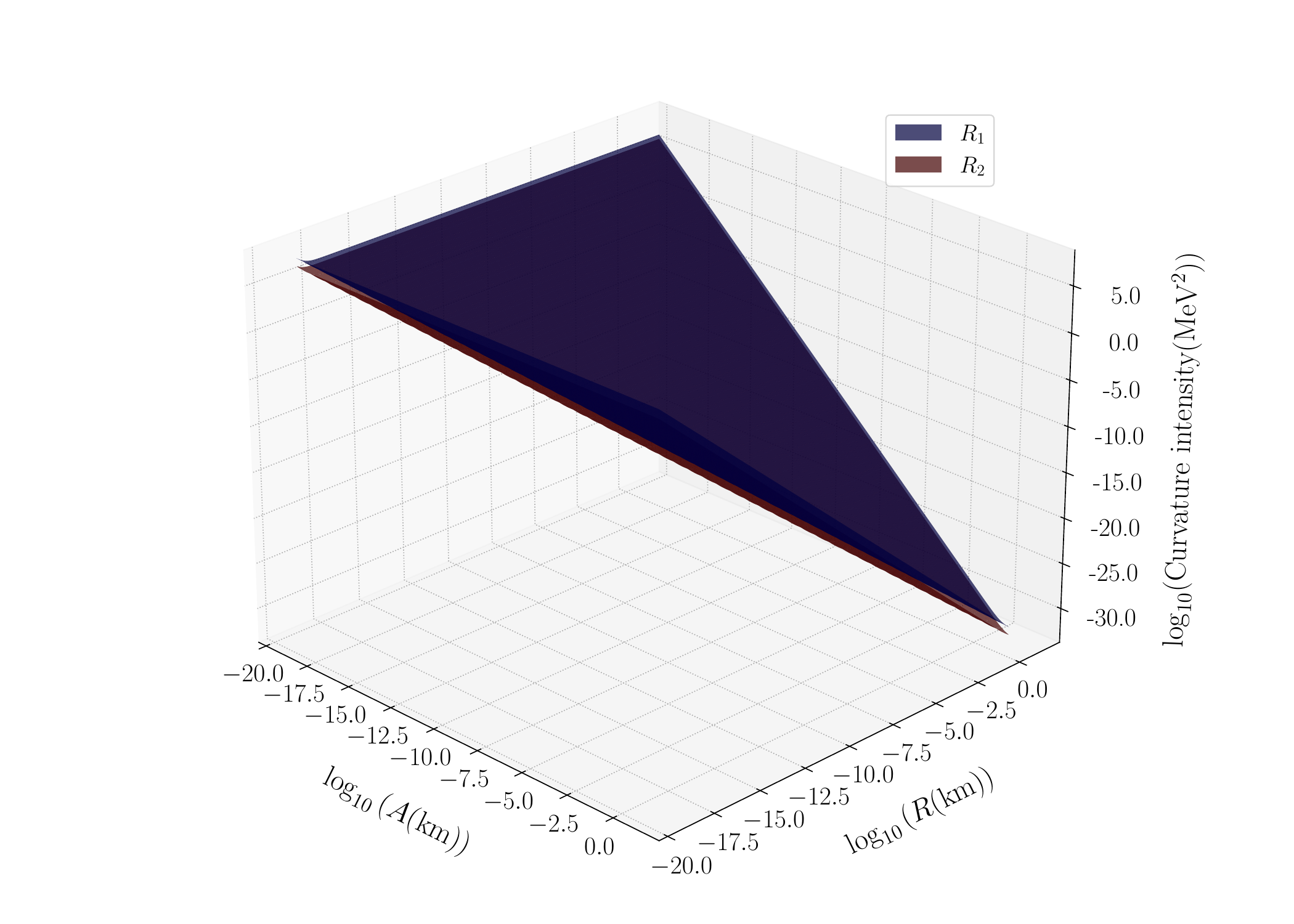}
    \caption{Values of $R_1$ and $R_2$ against parameters $R$ and $A$ at $r'=0$.}
    \label{fig:curvatures3DReA}
\end{figure}
}

\newpage

{\color{black}
We can see quantitatively the value of $\Delta V$ for some examples of stars fitted to the Tolman IV model \footnote{Considering the parameters to be the same as \cite{Zamperlini:2025nly}, which $\mu = m_\pi \approx 135\ \mathrm{MeV}$  and $\lambda=4450\ \mathrm{MeV}$ for interacting particles with mass $m_\Phi=m_p\approx939\ \mathrm{MeV}$.}, in the plot of \autoref{fig:DeltaV-5estrelas-TolmanIV}. In it, we see that objects with $1$,$5<r_\star/r_s<3 r_s$ can weaken the interaction for shorter ranges ($<3\ \mathrm{fm}$), in the region of the original potential well, with this range being smaller the greater the compactness. On the other hand, objects with $r_\star>3r_s$ have a decrease in potential energy for all ranges, enhancing the attractive character of the original interaction, in the order of $10^{-33}$--$10^{-34}\ \mathrm{MeV}$.

\begin{figure}[H]
    \centering
    \includegraphics[width=\linewidth]{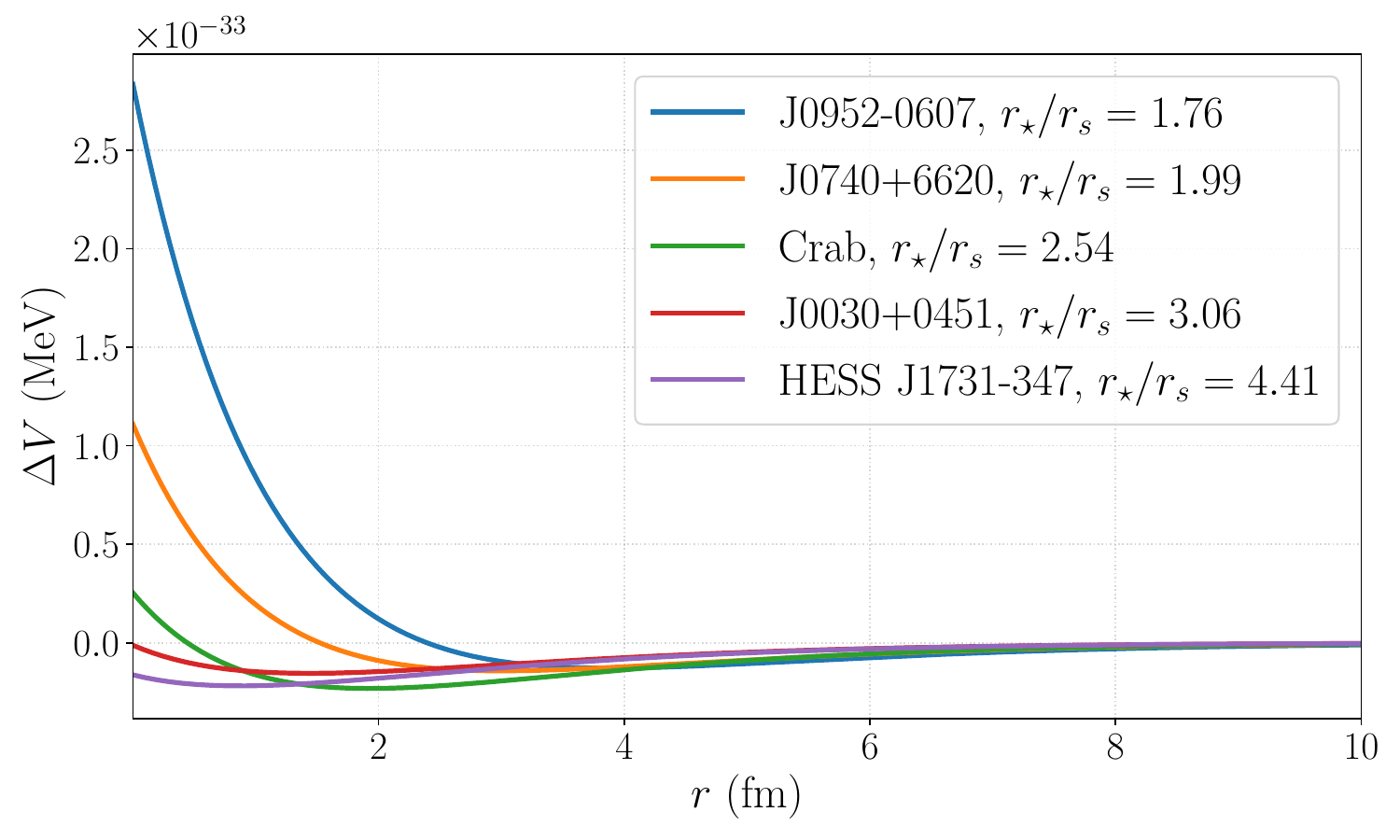}
    \caption{\textcolor{black}{Radial Yukawa potential energy shift due to spacetime curvature corrections in the Tolman-IV model, fitted to five compact objects.}}
    \label{fig:DeltaV-5estrelas-TolmanIV}
\end{figure}
}

{\color{black}
Considering the change on the local effective potential given by:
\begin{equation}
    V_{\mathrm{eff}}(g_{\mu\nu},r) = \frac{\ell(\ell+1)}{2m_\Phi \,r^2} + V(g_{\mu\nu},r) \quad,
\end{equation}
where $\ell$ is the angular momentum quantum number, we can adjust the data for various astrophysical objects and estimate the curvature-induced energy shift $\Delta V$. \autoref{tab:deltaV-5objects} shows that, for five objects, the energy shift of the potential at the minimum of the flat effective potential $V_{\mathrm{eff}}(\eta_{\mu\nu},r)$ \footnote{\textcolor{black}{The minimum is calculated by finding $\frac{d}{dr}V_{\mathrm{eff}}(\eta_{\mu\nu},r,\ell=1)|_{r=r_{\mathrm{min}}}=0$, which, for the parameters considered, gives $r_{\mathrm{min}}\approx1.173\,\mathrm{fm}$.}}  is on the order of $10^{-34}\ \mathrm{MeV}$.

\begin{table}[H]
    \centering
    \begin{tabular}{ccccc}
    \toprule[1.5pt]
        Object & ~~$M\ [\Msun]$~~ & ~~$r_\star \ [\mathrm{km}]$~~ & {\color{black}~~$r_\star/r_s$~~} & ~~$\Delta V(r_\mathrm{min}) \ [\mathrm{MeV}]$~~ \\
        \midrule[1.5pt]
        PSR J0952–0607 & 2.35 & 12.25 & {\color{black}1.764} &$6.54\times10^{-34}$ \\
        PSR J0740+6620 & 2.1 & 12.32 & {\color{black}1.986} & $1.15\times10^{-34}$ \\
        Crab Pulsar & 1.4 & 10.5 & {\color{black}2.539} & $-1.89\times10^{-34}$ \\
        PSR J0030+0451 & 1.44 & 13.02 & {\color{black}3.062} & $-1.52\times10^{-34}$ \\
        HESS J1731-347 & 0.8 & 10.42 & {\color{black}4.410} & $-2.14\times10^{-34}$ \\
    \bottomrule[1.5pt]
    \end{tabular}
        \caption{\textcolor{black}{$\Delta V$ (in $\mathrm{MeV}$) only at $r= r_{\mathrm{min}}$ for the system located at the center of the fluid sphere described by the Tolman-IV metric for \textcolor{black}{selected} different astrophysical objects \textcolor{black}{(with data extracted from the literature, which can be found in \cite{psrj07I-Miller:2021qha}, \cite{psrj07II-Riley:2021pdl}, \cite{psrj07III-Salmi:2022cgy}, \cite{HESS-Doroshenko:2022nwp}, \cite{psrj0030I-Riley:2019yda}, \cite{psrj0030II-Miller:2019cac}, \cite{crabpulsar-Bejger:2002ty}, and \cite{psrj0952-Romani:2022jhd})}, along with the mass, estimated radius considered for the calculation and compactness $r_\star/r_s$.}}
    \label{tab:deltaV-5objects}
\end{table}
}

\textcolor{black}{We stress that the extremely small magnitude of the energy shifts obtained for realistic neutron stars is not unexpected. Since, for compact stars the invariants relative to the curvature corrections in the Riemann normal coordinate expansion are many orders of magnitude smaller than the typical mass scale entering the Yukawa interaction. Consequently, the curvature-induced modification of the potential is parametrically suppressed by the hierarchy between the stellar curvature scale and the characteristic length scale of the interaction. The smallness therefore reflects the physical separation between nuclear and curvature scales in macroscopic compact stars, rather than a limitation of the formalism itself. We must remark that is possible to find astrophysical objects with configurations that present corrections of greater magnitude in determined regions. This analysis is left for future works.}

\newpage

\subsection{Tolman-VI Numerical Results}

The Tolman-VI solution (with $n=1/2$) lacks physical correspondence with realistic star models due to its fixed mass-radius relation, which does not align with observational data, and its singular center. However, it offers interesting aspects for comparison with the Tolman-IV solution.

\begin{figure}[H]
    \centering
    \includegraphics[width=\linewidth]{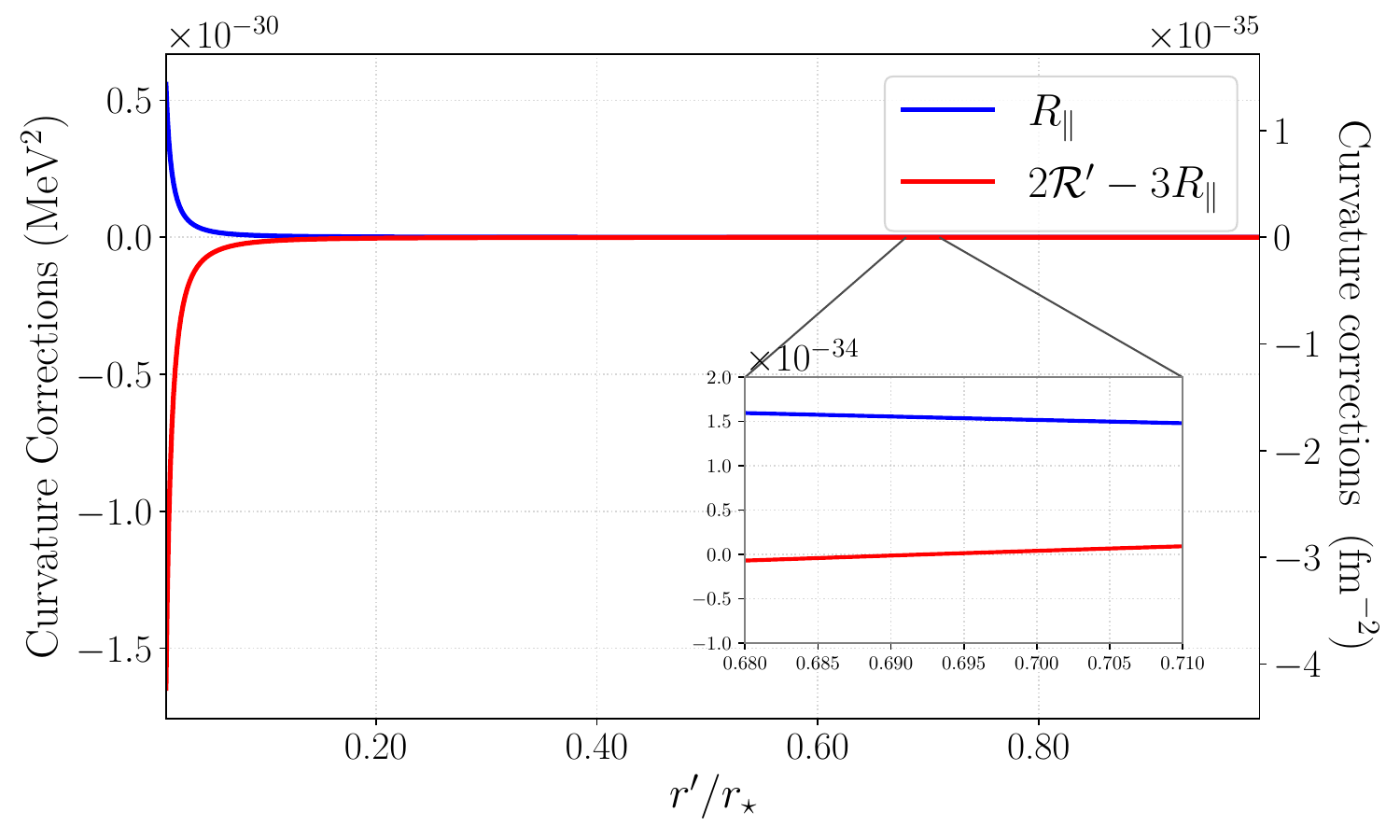}
    \caption{\textcolor{black}{Values of $\mathcal{R}'-3R_\parallel$ and $R_\parallel$ against $r'$ values for a fluid sphere of radius $r_\star=10\ \mathrm{km}$ and mass $M=1.45\Msun$ accordingly to the Tolman-VI solution.}}
    \label{fig:Tolman6-10km}
\end{figure}

For example, we can set a radius or a mass, and we can obtain all the curvature corrections inside the sphere. If we start with a star radius of $10\ \mathrm{km}$, then \autoref{eq:massradiustolman6} gives a mass of $1.45\ \Msun$ and the profile of the corrections can be seen on \autoref{fig:Tolman6-10km}, which also shows that the magnitude of the corrections is of the order of $10^{-30}\ \mathrm{MeV}^2$, comparable to the results from the Tolman-IV solution.

\textcolor{black}{Therefore, for Tolman-VI, the same order-of-magnitude smallness holds throughout the bulk of the fluid sphere when compared with the Tolman-IV results, with the clear exception of the unphysical singular center where curvature invariants diverge. Hence, excluding that pathological core, the curvature remains negligible at the Yukawa probe scale and the RNC-patch approximation used in our estimates remains applicable.}

{\color{black}
Furthermore, a similar analysis can be done for the change in curvature induced in the potential energy, as done for the Tolman-IV metric, that is, we can analyze the difference $\Delta V (r)$ when $r=r_\mathrm{min}$, using the same parameters. This way, we have the result of Figures \ref{fig:TO6-DeltaV-AgainstRprime-10kmstar} and \ref{fig:TO6-DeltaV-AgainstRprime-10kmstar-Zoomed}, when we use an object of $10\ \mathrm{km}$ radius. We observe that, despite being negligible quantitatively speaking, the behavior of the energy displacement brings aspects to be interpreted. It varies along the spherical fluid of Tolman-VI, and in the regions closest to the boundary at $r_\star$ the contribution of the curvature is negative, indicating a ``strengthening'' of the Yukawa interaction in these regions, but for a certain $r'/r_\star < r'_{\mathrm{crit}}/r_\star = 9/13 \approx 0.69$, the behavior of the curvature corrections alternates, inducing a positive correction (repulsive character) for a small local range of the potential $r <r_\mathrm{div}(r')$, found from \autoref{eq:potential-TolmanVI-rstar}:
\begin{equation}
    r_\mathrm{div}(r') = \frac{13\frac{r'}{r_\star}-9}{3+\frac{r'}{r_\star}}\cdot\frac{1}{\mu} \quad \text{for}\quad \frac{r'}{r_\star}>\frac{9}{13} ~~.
\end{equation}

These characteristics can be interpreted by considering the effective potential, potentially influencing the decrease of the effective potential well and a decrease of the potential barrier, altering the stability of the nuclear bond; extrapolating, this would indicate the breaking of this type of interaction in the vicinity of the singularity of the Tolman-VI metric.

\textcolor{black}{We reemphasize that the Tolman-VI solution possesses a central curvature singularity and a rigid mass–radius relation, which limits its astrophysical realism. Accordingly, results in the immediate vicinity of the center should not be interpreted as describing realistic neutron star cores. Instead, Tolman-VI is employed here as an analytic laboratory to explore how the formalism behaves in strongly curved interior geometries. Away from the singular center, the curvature corrections remain of the same suppressed order found in Tolman-IV, reinforcing the general conclusion regarding the smallness of effects in stellar-scale systems.}

\begin{figure}[H]
    \centering
    \includegraphics[width=\linewidth]{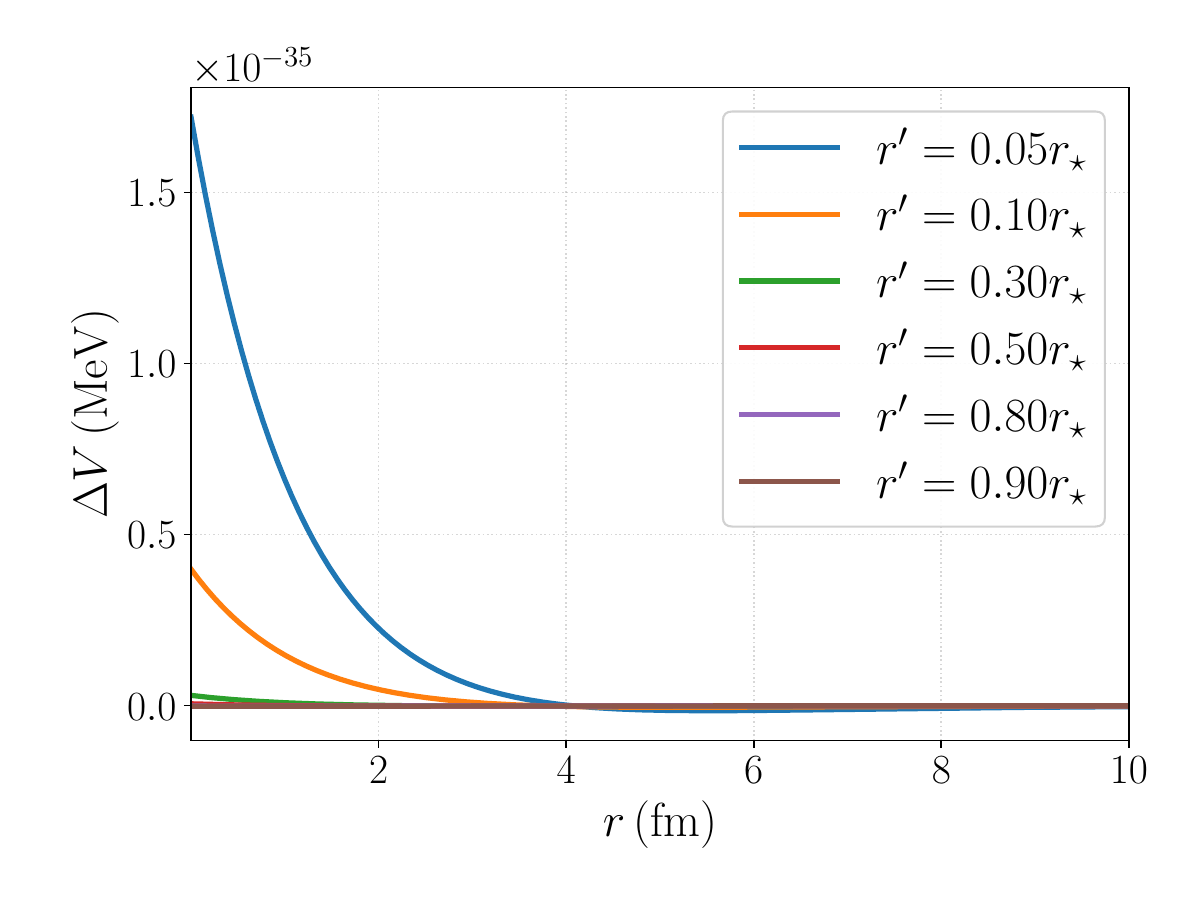}
    \caption{\textcolor{black}{Values of $\Delta V (r)$ for different values of $r'/r_\star$ for an object of $r_\star=10\ \mathrm{km}$ in the Tolman-VI metric.}}
    \label{fig:TO6-DeltaV-AgainstRprime-10kmstar}
\end{figure}

\begin{figure}[H]
    \centering
    \includegraphics[width=\linewidth]{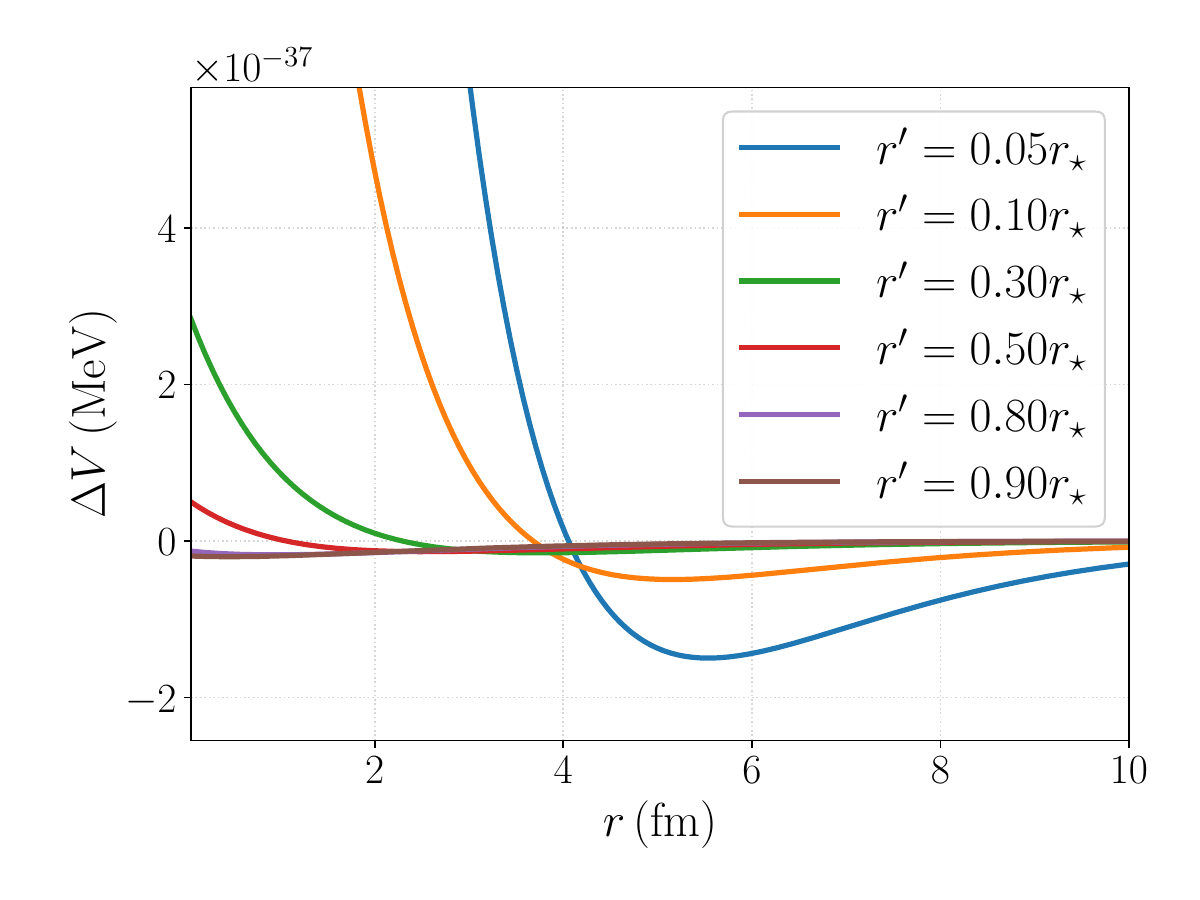}
    \caption{\textcolor{black}{Zoomed-in version of the $\Delta V(r)$ values for different values of $r'/r_\star$ for an object of $r_\star=10\ \mathrm{km}$ in the Tolman-VI metric.}}
    \label{fig:TO6-DeltaV-AgainstRprime-10kmstar-Zoomed}
\end{figure}

Following the analysis of compact objects, we can infer, at first glance, that gravitational effects would have a minimal impact on nuclear interactions, even in this high-pressure system, and likely also have a minimal effect on observables derived from an equation of state for nuclear matter. Thus, concluding that, within the scope of these phenomenological results, even in a simplified version, they indicate that the interaction of nuclear matter with primordial black holes is a more fertile ground for extracting non-negligible quantum effects in gravitation as fruits \cite{Zamperlini:2025nly}.

}

\section{Conclusions}\label{sec:conclusions}

In this work, we revisited the Yukawa potential derived from a $\phi^3$-like theory in curved spacetime, as presented in \cite{Zamperlini:2025nly}, and then applied the result to a generic spherically symmetric metric, with particular applications for the Tolman-IV and Tolman-VI metrics. In the Riemann normal coordinates, the corrections for the propagator have been determined, and the $\Phi\Phi$ potential for a one-boson exchange has been evaluated. By this procedure, it was possible to obtain the correction terms generated by the influence of the structure of spacetime. 

\textcolor{black}{The choice of the $\Phi\Phi$ interaction instead of the nucleon-nucleon interaction is due to technical aspects and may be viewed as a first approach to the problem. The curvature dependence of the corrected potential originates from the modification of the propagator in the local Riemann normal frame. While the scalar model produces a Yukawa-type central potential, a realistic nucleon–nucleon interaction would require Dirac fields and pion exchange, leading to the familiar one-pion-exchange (OPE) structure with central and tensor components. In such a formulation, curvature would modify both the scalar and tensorial sectors of the OPE potential through a similar geometric mechanism affecting the propagator, although with different spin-dependent coefficients. Therefore, the present analysis captures the geometric origin and order of magnitude of curvature corrections, while a full spinorial treatment is left for future works. Anyway, even considering the limitations of our approach, the analytical form of the potential that we obtain, with a correct set of parameters, generated results for the corrections that we believe that are of the order of the magnitude of the ones that we would find if Dirac fields had been considered in the formulation of the system.}

{\color{black} Applying the results for an expansion around a point $x'$ for a spherically symmetric metric, we defined local curvature corrections contributions in the parallel and transversal directions towards the metric center, simplifying the corrected potential; where anisotropy in the local potential exists if those two curvature terms are different. When evaluating for the Tolman metrics, the curvature contributions from the local Ricci tensor are the same for every direction, which we interpret as stemming from the condition of a perfect fluid in the energy-momentum tensor in Einstein equations, which has isotropic pressure; thus making the curvature-corrected potential radially symmetric in the local inertial frame.

The results for the Tolman IV solution depend on the compactness of the object considered. Depending on the regime of the values, the curvature corrections have different signs and then, different effects on the interacting potential. As for the numerical results, applying the solution to known astrophysical objects, which can represent compact stars, showed a correction of around $10^{-33}\ \mathrm{MeV}$ with the local frame at the regular center of solution IV. \textcolor{black}{In this sense, the neutron star estimates serve primarily as a quantitative benchmark for the formalism in realistic astrophysical environments.} For objects with $1.5<r_\star/r_s<3$, there are local regions (very short range) with positive energy shifts, indicating a weakening of the interacting potential, potentially in the region of the effective potential well, and negative energy shifts for the remaining range; whereas for $r_\star/r_s>3$, there are just negative ones, which we can interpret as a ``strengthening'' of the interaction due to spacetime curvature, but this would represent a deeper effective potential well but also a decreased effective potential barrier.

}
Additionally, we compare the results for the Tolman-VI solution, which is singular at the origin, with an infinite central pressure. Taking a choice suggested by Tolman, we calculate the curvature corrections and plot them for a fluid sphere with $10\ \mathrm{km}$ radius and $1.45$ solar masses. Such results showed a correction of the order $10^{-30}\ \mathrm{MeV}^2$ for most of the sphere radii, with a divergence at the origin. {\color{black} The numerical results show a very tiny energy shift due to curvature for the potential center located throughout the object radius, as example, at 5\% of the radius it is in the order of $10^{-35}\,\mathrm{MeV}$. Characteristic of this solution VI, there is some position of the system for which $r'<r'_{\mathrm{crit}}$, the corrections allow positive potential energy shift, possibly affecting the region of the effective potential well, decreasing its size, weakening the interaction, and for a larger range it shows a negative energy shift, possibly decreasing the effective potential barrier.
}

We also consider the comparison with \cite{Zamperlini:2025nly}, which focuses on the corrected potential outside the external event horizon of a Reissner-Nordström black hole. For a highly charged black hole ($Q=0.999Q_{\mathrm{lim}}$, where $Q_{\mathrm{lim}}$ is the maximum possible electric charge) with a mass of 3 solar masses, the corrections near the horizon are similar in magnitude to our results for the Tolman-IV and VI metrics, around $10^{-30}\ \mathrm{MeV}^2$. This fact is essentially due to the size of the considered astrophysical object. The results obtained suggest that these corrections could have significant implications for primordial black holes \cite{CarrEarlyBigBangBlackHoles:1974nx}, with event horizons located at very small distances from the center, providing larger values for the corrections than for the interiors of compact objects, and would be important in the study of the early Universe, \textcolor{black}{or even to extract some observable between the interaction of nuclear matter with this dark matter candidate \cite{Carr:2020-PBHasDarkMatter2}.}

Future research could explore more realistic relativistic star models. This includes numerically solving the Tolman-Oppenheimer-Volkoff (TOV) equations \cite{Oppenheimer:1939ne} with appropriate equations of state and investigating potential implications for these models using multi-messenger observations\textcolor{black}{; nuclear dynamics were studied in this context, for example in \cite{ETL-Maselli:2020uol}, \cite{ETL-Sabatucci:2023hpa}, \cite{ETL-Tiwari:2023tkj}, and \cite{ETL-Rose:2023uui}}.

Studies of the kind proposed in this paper are very important for investigating the behavior of quantum systems within the framework of general relativity, aiming to further elucidate the interplay between quantum fields and curved spacetime in the pursuit of a deeper description of nature.

\section{Acknowledgments}

We thank the Coordenação de Aperfeiçoamento de Pessoal de Nível Superior (CAPES), process number 88887.655373/2021-00, and Conselho Nacional de Desenvolvimento Científico e Tecnológico (CNPq) for the financial support, process number 312414/2021-8.

\bibliographystyle{ieeetr}
\bibliography{ref.bib}

\end{document}